\documentclass[sigconf]{acmart}

\AtBeginDocument{%
  \providecommand\BibTeX{{%
    \normalfont B\kern-0.5em{\scshape i\kern-0.25em b}\kern-0.8em\TeX}}}

\copyrightyear{2020}
\acmYear{2020}
\setcopyright{acmcopyright}\acmConference[IVA '20]{IVA '20: Proceedings of the 20th ACM International Conference on Intelligent Virtual Agents}{October 19--23, 2020}{Virtual Event, Scotland Uk}
\acmBooktitle{IVA '20: Proceedings of the 20th ACM International Conference on Intelligent Virtual Agents (IVA '20), October 19--23, 2020, Virtual Event, Scotland Uk}
\acmPrice{15.00}
\acmDOI{10.1145/3383652.3423900}
\acmISBN{978-1-4503-7586-3/20/09}

\usepackage{subfig}
\begin{document}

\title[A Virtual Conversational Agent for Teens with Autism]{A Virtual Conversational Agent for Teens with Autism Spectrum Disorder: Experimental Results and Design Lessons}

\author{Mohammad Rafayet Ali$^1$, Seyedeh Zahra Razavi$^1$, Raina Langevin$^2$, Abdullah Al Mamun$^1$}
\affiliation{%
  \institution{$^1$University of Rochester, $^2$University of Washington}
%   \city{Rochester, Seattle}
%   \state{New York}
  }
\email{{mali7,srazavi,amamun}@cs.rochester.edu}
\email{ rlangevi@uw.edu}

\author{Benjamin Kane$^1$, Reza Rawassizadeh$^2$, Lenhart K Schubert$^1$, Ehsan Hoque$^1$ }
\affiliation{%
  \institution{$^1$University of Rochester, $^2$Metropolitan College, Boston University}
%   \city{Rochester}
%   \state{New York}
  }
\email{{bkane2,schubert,mehoque}@cs.rochester.edu}
\email{rrawassizadeh@acm.org}

\renewcommand{\shortauthors}{Ali, et al.}

\begin{abstract}
  We present the design of an online social skills development interface for teenagers with autism spectrum disorder (ASD). The interface is intended to enable private conversation practice anywhere, anytime using a web-browser. Users converse informally with a virtual agent, receiving feedback on nonverbal cues in real-time, and summary feedback. The prototype was developed in consultation with an expert UX designer, two psychologists, and a pediatrician. Using the data from 47 individuals, feedback and dialogue generation were automated using a hidden Markov model and a schema-driven dialogue manager capable of handling multi-topic conversations. We conducted a study with nine high-functioning ASD teenagers. Through a thematic analysis of post-experiment interviews, identified several key design considerations, notably: 1) Users should be fully briefed at the outset about the purpose and limitations of the system, to avoid unrealistic expectations. 2) An interface should incorporate positive acknowledgment of behavior change. 3) Realistic appearance of a virtual agent and responsiveness are important in engaging users.  4) Conversation personalization, for instance in prompting laconic users for more input and reciprocal questions, would help the teenagers engage for longer terms and increase the system's utility. 
\end{abstract}




% <concept_id>10003120.10003123</concept_id>
% <concept_desc>Human-centered computing~Interaction design</concept_desc>
% <concept_significance>300</concept_significance>

% \ccsdesc[300]{Human-centered computing~Interaction design}


\maketitle

\section{Introduction}
Autism spectrum disorder (ASD) is a developmental disorder which affects one in 59 individuals in the US alone \cite{baio2018prevalence}. Almost all the individuals with ASD show deficits in nonverbal communication \cite{georgescu2014use}. 
% They often fail to make normative gestures, eye contact, and smile to make their verbal communication compelling. Thus, they often face difficulty in expressing their feelings and act out their frustrations through vocal outbursts \cite{maskey2013emotional,simonoff2008psychiatric,kasari2013assessing}. 
Current practices for improving the communication skills deficit involve therapy sessions with behavior experts. There is a significant shortage of behavioral experts resulting in therapy time being limited and inaccessible. A computer-mediated online social skills intervention has the potential to enable individuals with ASD to practice conversation frequently with a standardized and repeatable stimulus. Additionally, computer-mediated tools can enable a therapist to serve more individuals by monitoring their progress remotely while continuing their weekly face-to-face therapy. As an added benefit, such technology will enable remote therapy which is particularly useful when social distancing is recommended due to virus outbreaks (e.g., COVID-19). As we write this paper, we face a global pandemic that forced 91,000 U.S. public and private schools to close and affected more than 50 million school students \footnote{https://www.educationnext.org/covid-19-closed-schools-when-should-they-reopen-coronavirus} including people on autism spectrum. The sudden lockdown and extreme mitigation did not allow the therapists to prepare, leaving the individuals who are in need of behavioral therapy to look for online solutions. We believe online technologies such as LISSA has the potential to augment the traditional therapy for social skills development.

\begin{figure}
    \centering
    \includegraphics[width=8cm ]{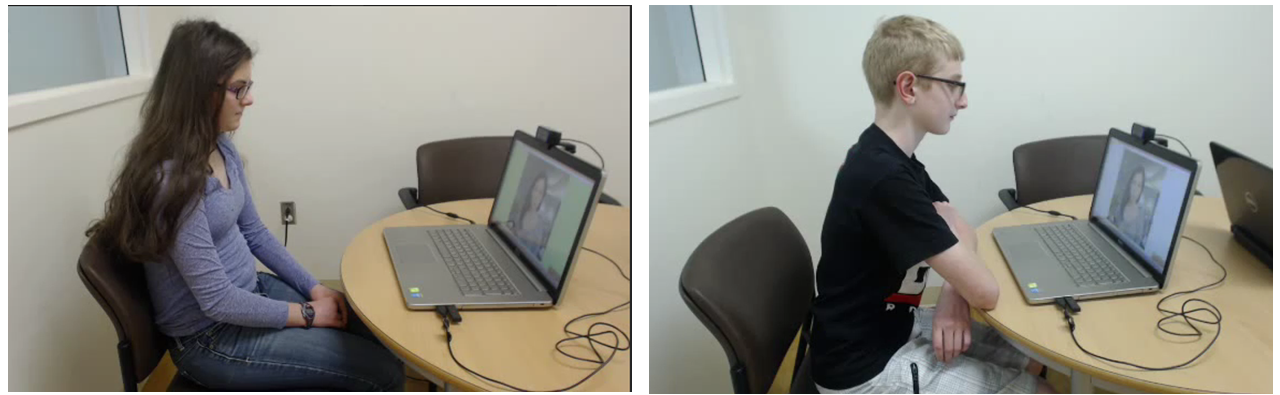}
    \caption{Teenagers with Autism Spectrum Disorder interacting with LISSA.}
    \label{fig:kids}
    \vspace{-4pt}
\end{figure}

% Although computers have many advantages, social skills intervention for individuals with ASD is very challenging. Dr. Stephen Shore, a professor of special education famously said, \textit{"If you've met one person with autism, you've met one person with autism"}. This indicates that a single solution does not exist, and if we want to design any intervention, it needs to be customized for individual needs. Thus, we need the active participation of individuals with ASD in the design process, so that the design will be appropriate both in typical cases and in terms of individual variation. Because of the deficit in communication skills a participatory design is often hard to achieve.   

In this paper, we review the design of an online social skills development program, LISSA - Live Interactive Social Skills Assistance. We focus on design lessons learned, based on trials with teens with ASD, particularly on analysis of post-session interviews with participants. LISSA features a virtual agent capable of engaging the user in a multi-topic conversation and giving real-time feedback on the nonverbal cues, by analyzing the spoken language and the facial expressions of the user in real-time. In addition, LISSA provides a summary feedback after the conversation. The motivating idea is that users can interact with LISSA repeatedly, and potentially learn and observe their skills improvement from the summary feedback in multiple conversation sessions, held in a private and safe environment. The design of the interface was guided by an expert UX designer, psychologists, and a pediatrician. The initial design of the interface was evaluated in a Wizard of Oz setting with 47 participants. The feedback system and the dialogue manager were then automated using the data. Subsequently, we focus on adapting the system design to teenagers with ASD to inform further development. The study, whose lessons we explore here, involved nine teenagers with ASD. Figure \ref{fig:kids} shows two teenagers with ASD interacting with LISSA in the lab. The teens with ASD interacted with LISSA and were interviewed about the experience. Through a thematic analysis of the interview transcripts we identified several key design guidelines, 1) Users should be fully briefed about the purpose of an interface, and its capabilities and limitations.  2) An interface should incorporate positive acknowledgment of behavior change. 3) Realistic appearance of a virtual agent and responsiveness are significant factors in engaging users.  4) Conversation personalization, e.g., user-sensitive turn-taking and prompting laconic users, would help the teenagers engage for longer terms, and thus have them benefit from the interaction.

Our paper makes the following contributions -
\begin{itemize}
    \item We motivate and explain the design of the LISSA system, and describe the automation of the system that makes it suitable for private, ubiquitous use. 
    \item Based on post-session interviews, we discuss key design guidelines that we identify as important for future interface design for individuals with ASD. 
\end{itemize}

% This paper is an expanded version of a conference paper \cite{razavi2016lissa} published in International Conference on Intelligent Virtual Agents (IVA 2016). The past conference paper introduces the LISSA system and the dialogue module. In this paper, we expand on the lessons learned from user studies as well as the system as a whole.  
\section{Related Work}
Individuals with autism spectrum disorder (ASD) are characterized in varying degrees by their social interaction, difficulties in verbal and nonverbal communication, and repetitive behaviors \cite{ozonoff2010prospective}.
% High-functioning ASD individuals are those diagnosed with autism but functioning cognitively at a relatively high level (e.g., IQ greater than 70) \cite{carpenter2009asperger,sanders2009qualitative}. However, high functioning ASD individuals may still demonstrate deficits in communication, emotion recognition, and social interaction \cite{sanders2009qualitative}. 
The existing treatments for high functioning ASD do not address the condition as a whole, rather they focus on individual symptoms. Thus there is no single intervention for such individuals. In this section, we discuss the existing computer-based social skills interventions for individuals with ASD. Additionally, we discuss the work that employed virtual agents for social skills training. 

\subsection{Social Skills Intervention for ASD}
In the past, interventions that focused on children and teenagers with ASD touched a variety of technologies, including virtual agents. 
% Gal et al. \cite{gal2016using} developed a tabletop application as an intervention for improving social skills among children with high functioning ASD. The application allowed two to four children to collaborate and create a narrative of a story for a given scenario. A study with 14 high functioning ASD children over a 3-week time period revealed that children were more likely to initiate positive social interaction.
Tartaro and Cassell \cite{tartaro2008playing} developed a virtual peer to engage children with ASD in a collaborative storytelling task. The results showed that children with autism approach the collaborative activities with virtual peer with excitement and they get engaged and improve in contingent discourse. These works prove the applicability of interactive technologies in social skills training for individuals with ASD. Feedback on nonverbal and verbal behavior was also applied as communication skills training. % Boyd et al. \cite{boyd2016saywat} designed SayWAT, a wearable assistive technology for individuals with autism, which provides real-time feedback on prosody during face-to-face interaction. They demonstrated that their tool could detect atypical prosody and deliver feedback in real time without disruption to the conversation. A few work focused on generating design guidelines. Hayes et al. \cite{hayes2010interactive} have provided three prototype systems addressing the design challenges with the use of large group displays, mobile personal devices, and personal recording technologies. Through a qualitative study with 13 children, they presented design guidance for visual support. 
% \subsection{Virtual Agent-Based Social Skills Training}
Foster et al. \cite{Foster2010} designed ECHOES as a multimodal learning environment intended for children with ASD. The system features a virtual agent that engages a child in a collaborative learning activity and provides feedback based on sensed features including gaze direction and gesture. In a subsequent work with ECHOES, Bernardini et al. \cite{Bernardini2013} designed a virtual agent as a credible social partner for children with ASD, that engages them in interactive learning activities. Milne et al. \cite{Milne2009} designed a virtual agent as an educational tool for children with ASD, which helps improve their conversational skills and ability to deal with bullying. In a study with ten participants, the authors showed that participants who received the intervention had gained higher conversational skills and more knowledge about bullying than the control participants. Mower et al. \cite{mower2011rachel} developed Rachel, an embodied conversational agent designed to elicit and analyze naturalistic interactions. This tool was designed for children with autism to encourage their affective and social behavior. Boujarwah et al. \cite{boujarwah2012socially} presented a tool to enable non-expert humans to generate conversational scenarios, which can be used to teach children with ASD, appropriate behaviors in different social scenarios. DeVault et al. \cite{devault2014simsensei} developed SimSensei - a virtual agent in the context of the healthcare decision support system. The goal of this system is different than our work as it aims to identify psychological distress indicators through a conversation with a patient in which the patient feels comfortable sharing information. This system has both nonverbal sensing and a dialogue manager. The dialogue manager uses four classifiers to categorize the users' speech, and hence to generate a relevant response. Hopkins et al. \cite{hopkins2011avatar} developed a social skills training system for children with ASD. In a study with 49 individuals, the authors demonstrated that the program helped the participants improve their emotion recognition ability and social interactions. 
Razavi et al. \cite{razavi2016lissa} developed a conversational agent capable of conversing with teenagers with ASD. The authors employed a script-like schema for guiding the dialogues, and generated appropriate responses using hierarchical pattern transductions. 

\subsection{Dialogue Management}
Smartphone-based conversational agents can be impressive in their question-answering tasks but have been found to provide incomplete and inconsistent responses in areas like mental health, interpersonal violence, and physical health \cite{miner2016smartphone}. Woebot \cite{fitzpatrick2017delivering}, a chat-bot recently developed at Stanford for delivering CBT to young adults with depression and anxiety, was able to significantly reduce symptoms of depression; however, its inability to hold a natural conversation was reported as the main issue users complained about. Popular methods in dialogue managements include frame-based methods such as \cite{fitzpatrick2017delivering} and \cite{razavi2017managing} where the user's responses are used to fill frame slots and to perform a task according to the contents of the frames. Another method, employed in SimCoach \cite{morbini2012mixed} and ASST \cite{tanaka2017embodied} controls dialogue via transitions between predetermined states based on the user's input. While the method shows good performance in limited dialogues, scalability remains an issue for more open-ended conversation. Some other end-to-end systems achieve meaningful single-turn responses through data-driven methods (e.g., \cite{Serban2016}), but cannot conduct a coherent longer dialogue; also, data-driven methods are susceptible to bias and privacy issues, among others \cite{Henderson2017}.

Taken together, prior studies in computer-based social skills training and dialogue management suggest that dialogue-based systems hold considerable promise for helping individuals with ASD (or certain others with special needs). In our work, we have focused on improving users' nonverbal behavior in communication through real-time feedback. Through participatory studies, and post-session interviews, we have identified significant design guidelines for such virtual agent based intervention. 

\section{System Design}
\begin{figure}[t]
     \centering
     \subfloat[LISSA Interface. The virtual agent engages the users in a conversation and provides real-time feedback by flashing the icons at the bottom.]{\includegraphics[width=0.3\textwidth ]{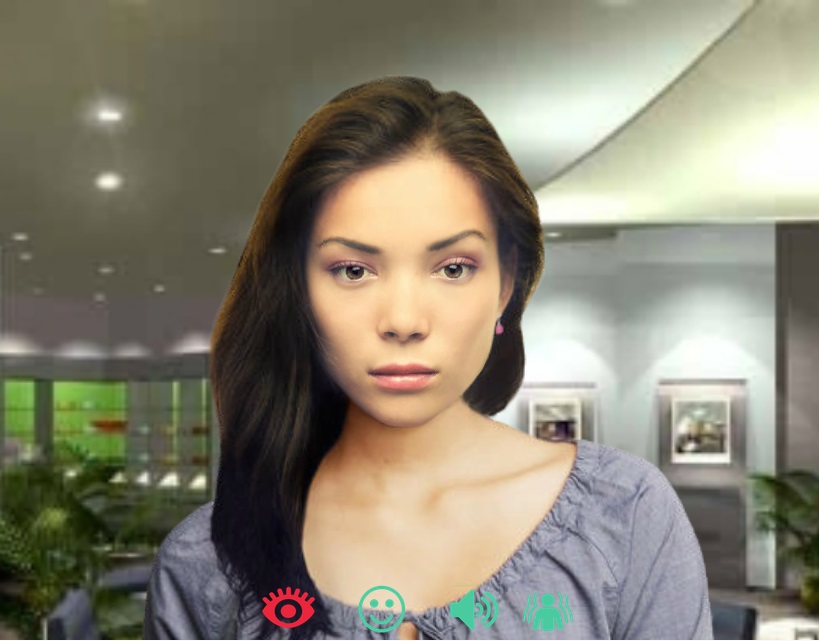}\label{fig:lissa}}\hspace{5mm}
     \subfloat[Post conversation summary feedback. Reminders: how often the users received feedback, Best Streak: how long the icons were green, Response Lag: how much time (on average) it took to turn a red icon green.]{\includegraphics[width=0.4\textwidth ]{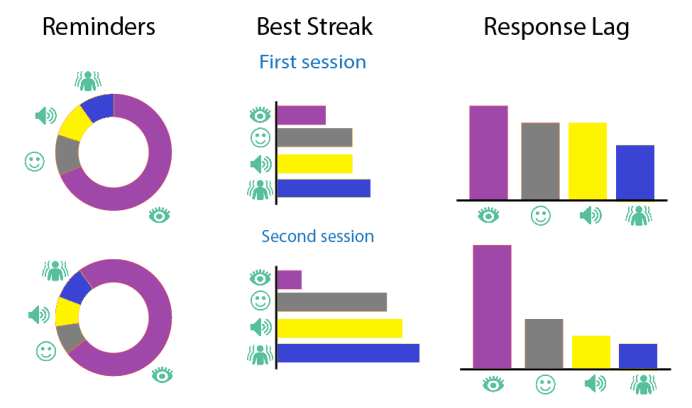} \label{fig:post-feedback}}
     \caption{LISSA system and post conversation feedback interface.}
     \label{fig:interface}
     \vspace{-8pt}
\end{figure}

In this section, we first describe the LISSA interface and the initial data collection process using a Wizard-of-Oz prototype. Then we describe the feedback and dialogue generation of the system. The automation of the system, prepared the way for the study with ASD teens, on which we focused later on.

\subsection{Interface}
Our design team includes two psychologists, a psychiatrist, and a user experience designer. The interface consists of two major parts: a conversation interface, and a post-conversation feedback interface. The conversation interface features a virtual agent (see Figure \ref{fig:lissa}), able to hold conversations with users, and provide real-time feedback on their nonverbal behavior. The feedback is presented through four flashing icons placed at the bottom of the interface. The four icons represent eye-contact, smile, speaking volume, and body movement. The icons are green by default but turn to flashing red as a prompt to the user to adjust their behavior. We kept the simple red-green color scheme to reduce the cognitive load on the user, since real-time feedback can be distracting \cite{Heimberg2002}. The persona of the virtual agent was selected from the recommendation of the pediatric psychiatrist of our team. The four nonverbal cues were selected because of their known applicability in improving social skills \cite{Feldstein2014}. The post-conversation feedback interface (see Figure \ref{fig:post-feedback}) shows a summary of the feedback provided during the conversation. This interface shows how many times the user received feedback (Reminder), how long the user kept the icons green (Best Streak), and how much time the user took to adjust nonverbal behavior (Response Lag). In the past, similar interface has been used for communication skills training \cite{Ali2015, razavi2016lissa}.

\subsection{Data Collection}
% \subsection{Wizard of Oz Prototype}
We first developed a Wizard of Oz prototype of the interface. There were two human operators for driving the system. One was responsible for the dialogue management and the other was responsible for giving feedback by flashing icons. Both operators were able to monitor the user remotely and control the interface through a web interface. This design allowed us to collect data from users and learn about their user experience without as yet applying machine learning techniques. Subsequently, we used the collected data to train machine learning models for automation. 

% To collect data and assess the viability of the interface we conducted a randomized control study with 47 college students and 8 female research assistants in the context of speed-dating. Speed-dating is gaining increasing popularity among researchers as a tool for studying social and communication skills \cite{Finkel2007}. The study results revealed that participants who used the LISSA interface improved their eye-contact and head nods \cite{Ali2015}.

We collected 46 videos of 23 neurotypical young adults (age between 18 and 23) interacting with LISSA. The videos were captured through a camera attached on top of the computer monitor. The Wizard of Oz operators were driving the LISSA program while we recorded the videos. We then employed six research assistants majoring in Psychology to label the collected videos. The psychologists in our team provided multiple training sessions to the research assistants. During the labeling the research assistants were additionally given a code book to follow. The code book includes rules for marking the moments where the participants should receive feedback. An example of these rule could be marking the moment as eye contact feedback when the participant is looking down. The research assistants watched each of the video recordings and marked those moments where the participants should receive red icon flash. The initial average inter-rater reliability was 0.42 (Krippendorff's alpha). We then decided to consider those moments for feedback labels where more than two research assistants marked it as a feedback moment due to the low overall agreement.    
% \begin{figure*}
%     \centering
%     \includegraphics[width=\textwidth]{figs/algo.png}
%     \caption{Feedback generation algorithm using HMM.}
%     \label{fig:algo}
%     \vspace{-5mm}
% \end{figure*}
\subsection{Automated System}
We extracted facial and prosodic features from the recorded videos, including head pose, smile, facial action units, volume, and voice pitch. For this, we used off-the-shelf software tools, namely OpenFace \cite{Baltru}, and Praat \cite{Boersma}. We then trained a hidden Markov model (HMM) to generate the flashing-icon feedback from the facial and prosodic features. The detailes and accuracy of this model can be found here \cite{ali2018aging}. The In the past, HMMs have proven successful in modeling human behaviors and actions \cite{Chung2008}. 
% Figure \ref{fig:algo} shows the training and feedback generation technique using the HMM. 

\begin{figure}
    \centering
    \includegraphics[width=8cm]{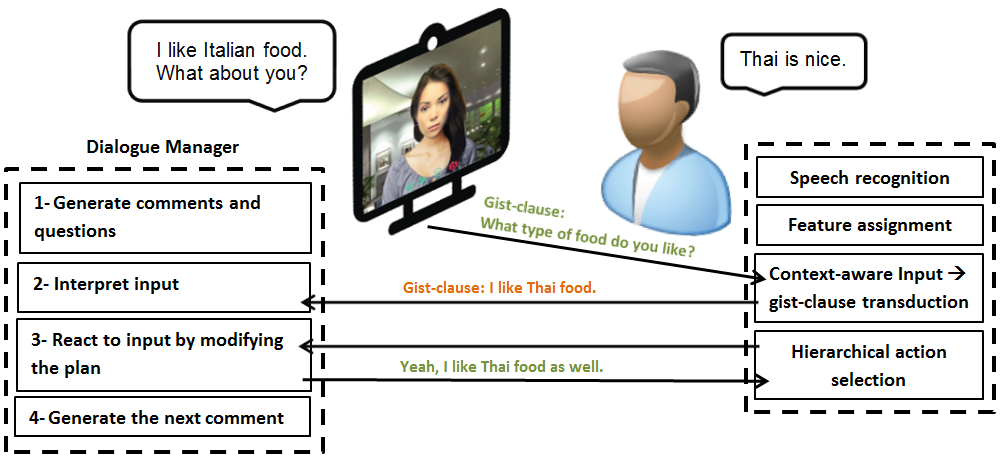}
    \caption{Overview of LISSA dialogue manager.}
    \label{fig:dialogue}
    \vspace{-4pt}
\end{figure}
\subsection{Dialogue Management}
In order for LISSA to conduct a conversation, we developed an automated dialogue manager capable of handling multiple topics. An outline of the functionality of the dialogue module is presented in Figure \ref{fig:dialogue}. LISSA leads the conversation by asking questions on different topics (often after a "personal" remark about the topic), and making relevant comments on the user's responses -- comments intended to show actual understanding of the user. The dialogue manager can also handle some questions asked by user. It continually updates the conversation plan, based on the user's responses.  At the top level, LISSA uses a structure called a schema, which contains a list of expected successive events in a dialogue; it allows for actions by both interlocutors and can be dynamically modified by user responses. The schemas are hierarchically structured, allowing LISSA to insert subschemas into the dialogue plan, helping to make the conversation more spontaneous. In order to capture users' inputs, we use the Nuance \cite{Nuance} speech recognizer. 

\begin{figure*}
    \centering
    \includegraphics[width=14cm]{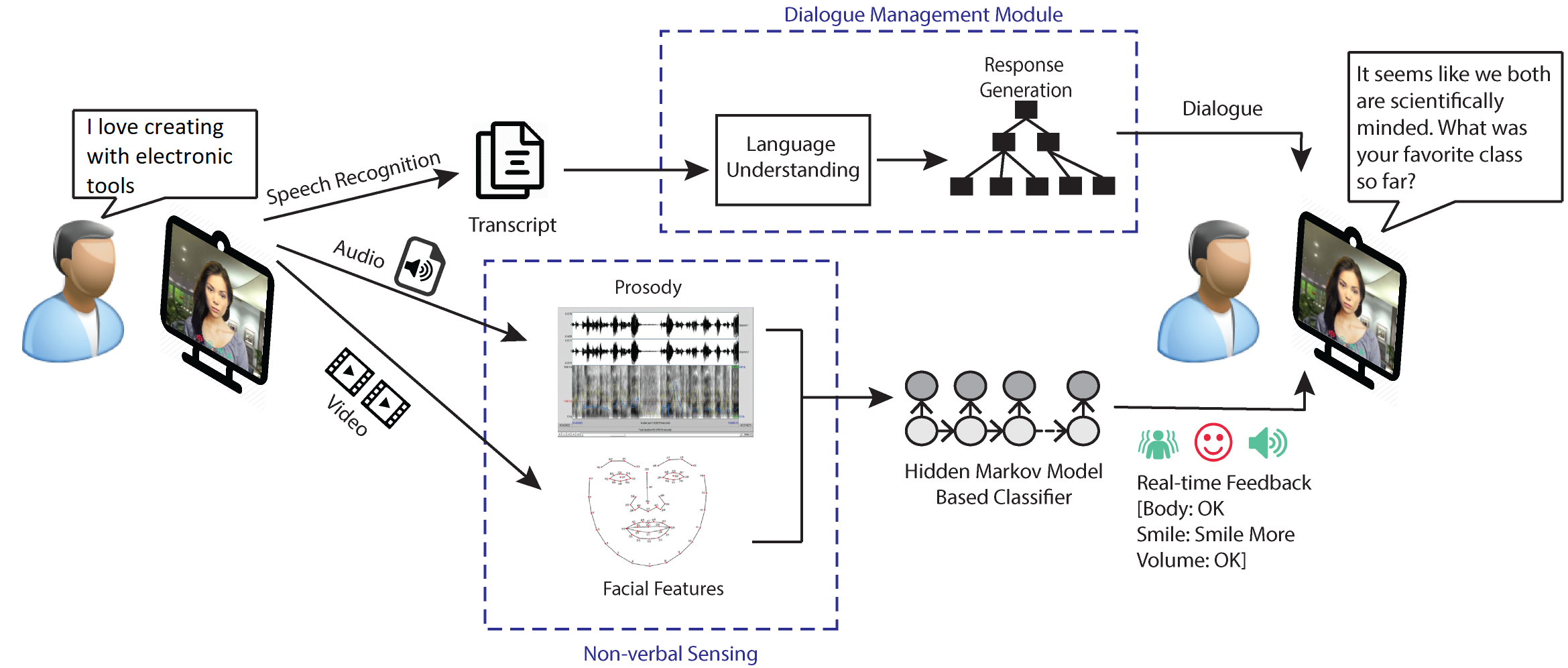}
    \caption{(From left to right) User having conversation with LISSA. System captures the speech, audio and video. Dialogue Management Module generates response. Nonverbal sensing extracts features and feeds to a hidden Markov Model based classifier, which produces real-time feedback in the form of flashing icon.}
    \label{fig:process}
    \vspace{-4pt}
\end{figure*}

When LISSA's dialogue manager receives input from the speech recognizer, it generates a high-level interpretation in the form of short, explicit, context-independent English sentences which we call "gist clauses". Gist clauses are extracted by applying several pattern transduction trees to the user's input, taking the current LISSA question as the context. In order to facilitate input matching, input words are automatically annotated with syntactic and semantic features before extraction of any gist clauses. Features are recursively attached to input words, such as GOODPRED for words like "happy", and ones like SOCIAL-SCIENCE and (by recursion) ACADEMIC-SUBJECT for "linguistics".  

After gist clauses have been derived from the user's input, they undergo a second stage of transduction, producing a reaction for LISSA to output. If the user's input answers a question by LISSA, the reaction will usually be a relevant reaction to that answer; or, if there is a question among the extracted gist clauses (typically at the end of a user's input), LISSA is likely to answer the question. Every gist clause obtained is stored in LISSA's memory so that LISSA won't ask for information already provided by the user in previous turns. Also, the collected gist-clauses could be used for future enhancements that allow inference during the conversation and reference to previous contributions of the participants. Figure \ref{fig:process} provides an overview of the functioning of the system. More details on the dialogue manager can be found in ~\cite{razavi2017managing}

\section{Study}
Over the past three years, we have been recruiting participants for our studies through the developmental/behavioral pediatric research center at the local medical center. Our experimental sessions have been conducted with nine teenage participants (i.e., between 13 and 17 years old) with high functioning ASD. There were one female and eight male participants. All participants were white and native English speakers. The pediatric research center allowed us to contact the participants who were already receiving behavioral therapy and expressed their willingness to participate in research studies. The diagnosis of the participants were also done by the research center. We made sure that our participants had the ability to read and able to have a conversation. 

The goal of this study has been to learn what aspects of LISSA are judged to be useful, and what adjustments need to be made in order to make LISSA a useful tool for iterative conversation training in the lives of teens with ASD anywhere. The sessions with the teens (whose parents were also invited) were scheduled on separate days. Each participant first interacted with LISSA for five minutes, then took a break for two minutes, followed by a second conversation with LISSA for another four minutes. As dialogue topics, we picked ones common in casual conversations such as "getting to know each other", "living in the current city", "crazy room" (aimed at eliciting imaginative responses), "city I want to move to in future", "free time", and "movies". During the conversations, the participants received real-time feedback through the flashing icons. After each conversation, the participants received post-session summary feedback. We conducted an interview with both the accompanying parent and the participant right after the LISSA session. The interview included survey questions on LISSA's usability and capacity for open-ended discussion. The interview was audio recorded and then transcribed by professional transcribers.
\section{Survey Results}

\begin{figure*}
  \includegraphics[width=14cm]{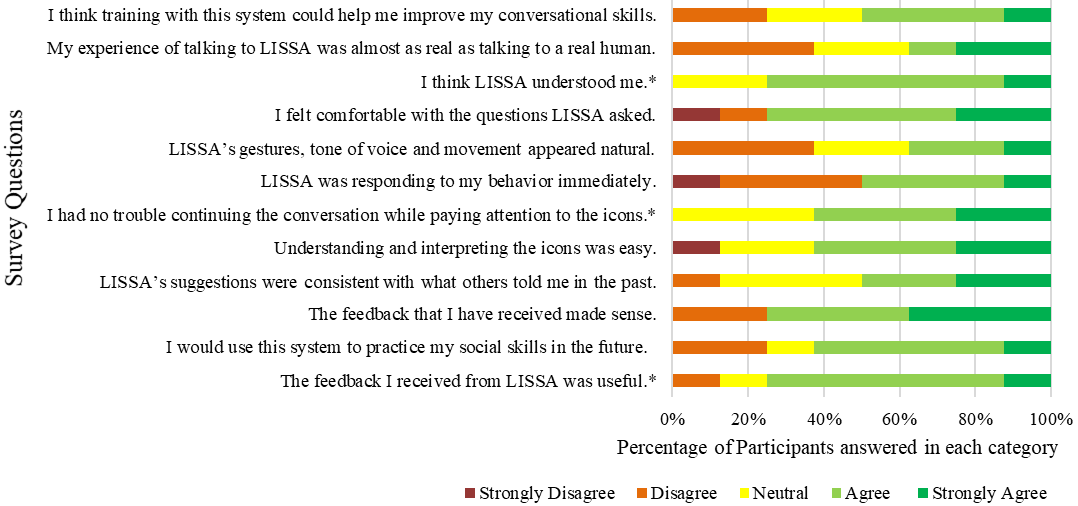}
  \caption{Survey questions and responses.}
  \label{fig:survey_results}
\end{figure*}

We presented 12 statements to the participants and asked them to specify their opinion ('strongly disagree' to 'strongly agree'). The questions were targeted to understand the usefulness of the feedback and the dialogues. Figure \ref{fig:survey_results} shows the specific questions and percentage of participants' answers in each category. The questions marked with a star (*) were answered significantly more $(p<0.05)$ with options 'agree' or 'strongly agree' compared to other options. We performed a single sampled non-parametric significance test \cite{mann1947} with Bonferroni correction \cite{Armstrong2014} against the option 'neutral'. Participants felt that they were being understood by LISSA.  Participants also expressed that they could continue the conversation and pay attention to the icons without any trouble. This indicates that real-time feedback might be applicable to the teenage population. Additionally, participants felt that the feedback they received from LISSA was useful. 
% During our interview session participants expanded on this perceived usefulness. The feedback was consistent and in accord with what their therapist said in the past. For instance, one participant had issues (i.e., slouching) with his posture and he received feedback through the 'body movement' icon. During the interview, the participant mentioned this and said that his parents often ask him to sit straight.
As can be seen in the figure \ref{fig:survey_results}, participants had mixed opinions about several questions, such as whether the conversational experience felt real, and whether LISSA's movements seemed natural. Three participants responded positively about the latter question, four responded negatively, and two were neutral. In the interview session, several mentioned that the lip movement and eye gaze were unnatural. Additionally, LISSA was not responding immediately. This was due to the fact that LISSA processes the dialogues and facial features in real-time and the processing takes place on a remote server. In our future versions of LISSA, we will make it more responsive by performing most of the computing locally. 
\section{Qualitative Analysis}
We performed a thematic analysis \cite{Guest2012} on the interview transcripts. In the past, thematic analysis was successfully used for user-centered design \cite{McCurdie2012} and rapid online interface prototyping \cite{Kinzie2002}.  Additionally, thematic analysis was used for identifying the design guidelines for developing computer and phone-based technology for children with ASD \cite{Mintz2013}.  In our analysis, three researchers performed thematic analysis and then the themes were merged to produce the final analysis report. As a basis for a qualitative thematic analysis of the interview transcripts, we considered individual interviews from the perspective of the following 14 labels: usefulness, perceptiveness, related systems, accuracy, familiarity, curiosity, realism, speed, appearance, improvements, social, multitasking, uncertainty, and adult identity. Five themes (closely related to some of these labels) stood out to us as relevant to summarizing the experiences of the participants. These are elaborated in the following. We believe it is at least as important to focus on weaknesses as on strengths, as a guide to further development.

\subsection{Utility for Practicing Conversation in Private}
Participants generally found LISSA useful for practicing conversations. Additionally, they thought that the program was not hard to navigate. One participant said, 
\begin{quote}
    \textit{"Wasn't that hard to use. It could definitely be used for somebody who really needs help with conversations or for somebody who is not really social or for somebody who is not really the kind of person to be talkative."}
\end{quote}
Participants liked the fact that the feedback was not coming from a human and they could use it in their private space without being observed. When we asked if they prefer human or a computer for giving feedback, some were ambivalent, some preferred the computer (\textit{"I would rather have that (LISSA) for feedback,"}), and some were skeptical about LISSA.

Some participants with favorable reactions added that the automatic facial feature detection and the accuracy of the feedback were the main reason for endorsing LISSA for conversational skills training. One participant said,
\begin{quote}
    \textit{"The fact that it was able to actually detect the facial features and everything being so accurate, I would consider that is good enough to actually train on."}
\end{quote}
The caregivers, as well, liked the fact that LISSA allows users to have a conversation with a virtual agent instead of (for example) a stranger online. They felt that LISSA was realistic and appreciated that it provided quite complex, and private interactions. 

\subsection{Self-awareness}
When conversing with LISSA, participants were often evaluating the experience in relation to real, social settings, for example, whether the feedback they received was truly appropriate. One participant noted that smiling broadly enough for LISSA to notice might be perceived as inappropriate in a public setting:
\begin{quote}
    \textit{"If they can make it so that she can respond a little quicker and that she is able to pick up the smile little better because I was smiling a little bit she just didn't think it was good enough and if I go too big then people are gonna think I am creepy."}
\end{quote}
Occasionally the self-aware evaluation of LISSA took the form of push-back against the presumption that they needed to improve their behavior, implicit in LISSA's feedback. One participant indicated that any inadequacy in their behavior with LISSA was due to the awkwardness of interacting with a virtual agent:
\begin{quote}
   \textit{"Well, I am actually social. I do make good eye contact with other people but it's just kinda awkward you know."} 
\end{quote}

From this it appears that some participants would need some time to become more familiar with LISSA. After multiple interactions they might well feel more comfortable with the system and be more accepting if the purpose of the system were more fully explained.

% \subsection{Realism of the virtual human}
% Participants focused quite intently on the realism and precision of the LISSA persona. There were debates on how realistic the eyes, lips, face, voice, and overall movement were. Participants suggested various improvements, such as more flowing speech, immediate responses, and faster blinks:
% \begin{quote}
%     \textit{"The other part I didn't like about her is because it took her a while to respond to my statements. So it was kind of confusing and irritating."}
% \end{quote}
% \begin{quote}
%      \textit{"I would say blinking eyes was definitely a little bit slow, because if I blink it looks almost instant. But for that it was half a second total, it seems quite slow."}
% \end{quote}
% \begin{quote}
%     \textit{"Nice sounding voice, nice face but the moving the lips thing was kind of a little creepy. Mainly because kind of it's little too computer-ish maybe."}
% \end{quote}
% When participants were asked about the usefulness of LISSA, their judgments were dependent on realism. For the other interview questions, participants had varying responses, but there always seemed to be an underlying fixation on realism. 
% \begin{quote}
%   \textit{"Well, it was good that she asked me what I liked to do. But since it wasn't really real it just seemed all awkward for me. I would make it better by making the quality High Definition and High quality better with talking to her for real."}  
% \end{quote}

\subsection{Multi-tasking and Feedback}
When they were asked about multitasking between the conversations with LISSA and looking at the icons, they wanted more prominent and bigger icons, that is easy to focus on. 
\begin{quote}
    \textit{"Maybe the icons at the bottom make it a like little bigger and maybe make it a little stand out a little bit more."}
\end{quote}
And when asked about if they preferred feedback from a computer or human, the participants had mixed opinions. For example, one participant said, 
\begin{quote}
   \textit{"The feedback I received from LISSA was useful. Well it was kind of a bit choppy and kind of pretty much prefer something a bit more realistic."} 
\end{quote}
Interpreting feedback through LISSA's behavior while conducting a conversation may be overwhelming for many participants. Perhaps if LISSA herself provided the feedback verbally during the conversations, rather than relying entirely on icons for feedback, it would feel more like a real interaction to users. Regardless of their preferences in feedback delivery, all participants expressed that the feedback was consistent with what other people had said to them in the past face-to-face training sessions. An important observation was that users broadly agreed on the need for positive feedback. While the flashing red icons noticeably indicated the need for improvement, the reversion to static green icons was not sufficiently noticeable as positive feedback. 

\subsection{Related Systems} 
At the end of the interviews, participants tended to compare LISSA to other conversational agents. They were often familiar with Siri and Alexa, and discussed the standard set by these virtual assistants in terms of speed and knowledgeability. They liked that LISSA could detect behavioral features, but hoped that LISSA could better understand and recognize them in the future. For example, a participant said, 
\begin{quote}
   \textit{"In its current state yes I might mess around with it some but I don't believe I would use it as an actual social skills training and jump into actual conversations just yet." } 
\end{quote}
It seemed that they had high expectations of LISSA which in some cases led to impatience during the conversation. One participant said they would feel more comfortable talking with systems like Siri, that they believe understands them. 
\begin{quote}
  \textit{"That's like my first time ever talking to a computer, well except for Siri that's different though. That one is fine."}  
\end{quote}

\subsection{The Desire for Adult Identity}
One teenager, after being urged by a caregiver to be honest, admitted that they probably wouldn't choose to interact with LISSA, unless perhaps LISSA ``popped up on their computer''. Part of the reason seemed to be that LISSA straddles the boundary between fantasy and reality.
\begin{quote}
    \textit{"I don't know if it was like a real person. Cause I do like fantasy things but I am also a kid who but when it comes to talking to people I like talking to real people."}
\end{quote}
These participants were ambivalent about the future use of LISSA because they were inclined to regard it as a conversational tool for children. One participant said their schedule was too busy, but others could benefit from LISSA.
\begin{quote}
    \textit{"Just so you know I am already 17 years old, I am growing up and some of these little kids' things I have outgrown but not all of them."}
\end{quote}
% In one conversation, a participant felt momentarily uncomfortable with LISSA's comment about a "crazy room," (asking them to speculate what kind of crazy room they would enjoy) and said they wanted to feel like an average adult: 
% \begin{quote}
%     \textit{"Well, pretty much just the crazy room cause well I wanna be what your average adult is. Basically responsible, kind, but also a bit unique."}
% \end{quote}
These comments again suggest the need to make LISSA's purpose clearer to users. It is not intended to be a surrogate human, but rather a tool for repetitive, private practice of conversational behavior. 
% Also, being sensitive in the choice of terminology is important, and perhaps asking for a creative response is inappropriate for some participants. 

\section{Discussion}

\subsection{Lessons about Interface Design for ASD Teens}
A majority of participants found that LISSA provided useful feedback. They also liked the fact that LISSA would allow them to converse in private. Our interviews helped shed light on how the interface design could be improved. The qualitative analysis of these interviews provide evidence for the soundness of our design so far and grounds for optimism about our further development plans. The analysis also allowed us to formulate several important guidelines for the design of LISSA-like interfaces for conversation practice. 

\subsubsection{Appropriate Prior Briefing of Users about LISSA's Purpose}
Our experience with users made clear that users' assessment of LISSA's behavior and potential utility for conversation practice depended very much on their expectations. They generally acknowledged that LISSA seemed to understand them and responded appropriately to inputs, yet was not genuinely human-like. The perceived shortcomings concerned LISSA's physical behavior, accuracy of perception of user behavior, and depth of knowledge. In part, this perception arose from comparisons with commercial systems such as Alexa and Siri, which have been optimized for smooth functioning in targeted information retrieval and other assistive functions. 

These reactions indicate the need for fuller preparation of users about LISSA's purpose: It is not a surrogate human, and it is not an app for access to useful knowledge or personal assistance. It is simply a tool for repeatedly practicing casual conversation for those who feel they could improve in that area. While LISSA has a range of verbal reactions to users depending on their particular inputs, and provides nonverbal feedback as a function of the user's behavior, the conversations are bound to be shallow, and to become more repetitive with multiple uses. Further, LISSA's physical behavior is not the focus; it is merely intended to be sufficiently human-like to make a casual conversation possible. All this should be made clear to potential users – along with a comment that there is no assumption that all users with ASD are lacking in the skills that LISSA is intended to help with. Such preparatory information could be provided both in advance of actual use of LISSA, and as part of LISSA's opening remarks (which already include \textit{"I might sound a bit choppy, but I am still able to have a conversation with you"}).

\subsubsection{Positive Acknowledgment of Behavior Change}
As noted in the previous section, the participants wanted to be made aware of positive changes. Perhaps flashing green could be used for behavioral improvements. Better yet, the virtual agent could say, for example, "You have good eye contact now". The efficacy of positive feedback and acknowledgment has been observed in past research \cite{Scott-VanZeeland2010,Kasari1988} and our experience further  confirms the desirability of positive acknowledgments for interventions aimed at social skills development.   

\subsubsection{Realistic appearance of a virtual agent and responsiveness}
Notwithstanding disclaimers about LISSA's physical behavior, the issue deserves further attention. A possible reaction to users' comments about insufficiently realistic smiling, eye blinking, and reaction speed might be to back away from the "uncanny valley" (e.g., Chattopadhyay et al. \cite{Chattopadhyay2016}) by using a more cartoon-like avatar. However, this would risk reducing LISSA to a toy in the eyes of potential users. Instead, we interpret the users' comments as urging further development of the avatar towards greater realism. Participants' comments suggest that the more life-like the character, the more likely they are to take it seriously. Realistic appearance of virtual characters has also been shown to be effective in other scenarios such as negotiation, tactical questioning etc. \cite{Kenny2007}. 
% The most important areas for improvement seem to be smiling and eye blinking. (Smiling is of course well-known to be very important in communication.) For example, smiling needs to be consistent with current feedback (one participant commented on co-occurrence of a smile by LISSA with negative feedback). Furthermore, smiles could be used directly to indicate improvements, or in support of positive verbal or icon feedback. 

\subsubsection{Conversation personalization would help the teenagers engage for longer terms.}
Some participants expressed enthusiasm about home-use of LISSA. However, they varied in their opinions on the choices of topics. For example, while the "crazy room" topic struck a chord with some (e.g., they would fill it with video games), others termed it childish. Participants thought that it would be useful if LISSA could talk about topics of their own choosing. For example, one participant was interested in computer programming and wanted to talk about it. The LISSA program, at its current stage, is designed for initiating conversational topics, and treating a specialized topic like computer programming seriously would be a major challenge. However, adding further mundane topics is feasible. Thus we could personalize interactions to a considerable degree by having LISSA choose topics dynamically, skipping those that the user seems indifferent to. Also, choices could be made sensitive to the user's age or maturity. Some of our newly developed topics pertain only to seniors, just as some of LISSA's topics for teens, such as bullying at school, pertain only to school-age users.

Another opportunity for personalization lies in the verbosity of the user. Our experiments showed that while some users provide expansive responses to LISSA, others respond tersely. As the goal of the system is to help users improve their communication skills, the system could gauge users' verbosity and provide helpful feedback where appropriate. For instance, LISSA might encourage laconic users elaborate their answers, or conversely, provide gentle suggestions about curtailing rambling or off-topic inputs. Assessing users' verbosity throughout the conversation may improve turn-taking behavior.  Users who tend towards longer answers should probably be allowed slightly longer silences before the turn is seized from them. The same applies to hesitant, slower speakers. 

% One of the most important observations we made about teens with ASD in comparison with (neurotypical) college students, independently of verbosity, was that the teens refrained from asking any reciprocal or other questions of LISSA (e.g., after telling LISSA about their favorite movie, asking "and what's your favorite movie?"). Whether this is due to less willingness to treat LISSA as human-like, or to limitations in social intuition, it is an area where verbal feedback by LISSA could be particularly useful; for instance LISSA might say, "This would be a good point to ask me about my favorite movie. Would you like to try?".

\subsection{Limitations and Future Work}

The current version of LISSA was not designed for immediate use in a randomized control intervention study. Rather, it is an exploratory system, which will enable a randomized control study after modification and enhancement based on the lessons learned from the trials with the teens with ASD. While our human-centric exploration adds important information on the design of interface and dialogue for a unique application such as autism, running a clinical trial remains part of our immediate future work. Additionally, in the future we will include color-blind friendly color palette for feedback.  

% We have used data from a different study with neurotypical individuals to train LISSA's feedback. We believe that there is a potential negative impact to the ASD teens participating on a highly exploratory study receiving feedback from human-wizards who were not trained therapists. On the other hand, the cognitive overload of the wizards giving live feedback through a computer mediated platform would have been further exacerbated having to deal with participants with social impairment. Given the huge individual differences and how ASD teens may react to such feedback, there were concerns about the consistency of such data. 

Our use of neurotypical individuals to train LISSA's feedback may seem nonoptimal, but we thought that using wizards not trained as therapists might have led to negative perceptions by ASD participants; furthermore, the wizards might have had difficulty with the cognitive load of managing the computer mediated platform, and providing live feedback to the ASD teens, taking account of the great individual differences in their reactions.

LISSA's dialogue manager was anticipating needs of teens with ASD, with advice from experts. It worked well, but the experiments have shown where improvements are most desirable, for example in topical adaptation to the user, inclusion of direct helpful hints in the verbal reactions to the user, and allowance for different turn-taking styles. Also, while the post-session interviews indicated that the users liked the appearance and voice of the avatar, they saw a need for improvements in the naturalness of the avatar's behavior. In our future work, we will design a customizable interface based on the knowledge we gathered through this study. 

Data collected from teenagers with ASD using future versions of the system will help us to further improve the sensitivity and responsiveness of the system. In the current system, the dialogue and the feedback modules are independent. It clearly would be useful to tie the nonverbal feedback to the dialogue content, and to supplement nonverbal feedback signals with direct verbal ones. In the future, we will make the feedback dialogue aware. 

\section{Conclusion}
In this paper, we described an interface capable of conducting a multi-topic conversation and provide feedback to help improve the user's overt behavior. The design benefited from the expertise of a pediatrician, psychologists, and a UX designer. We investigated further design desiderata through a study with nine teenagers with ASD. Using a thematic analysis we formulated several guidelines for improved interface design for teens with ASD. In future this knowledge will help guide interface design for this population as well as others with similar needs for social skills enhancement in casual dialogues. 
\section*{Acknowledgement}
This research was funded by the following grants NSF NRT-DESE Award 1449828, NSF IIS-1543758, DARPA subcontract  W911NF-15-1-0542-1. 

% We gratefully acknowledge the invaluable help and inspiration we received from the late Dr. Tristram Smith, without whom this work would not have been possible. 
The team is indebted to late Dr. Tristram Smith for his contributions to data collection for this project. 

\bibliographystyle{unsrt}
\bibliography{acmart}
\end{document}